\documentclass[preprint]{vgtc}               

\ifpdf
  \pdfoutput=1\relax                   
  \usepackage{graphicx}                
  \DeclareGraphicsExtensions{.pdf,.png,.jpg,.jpeg} 
\else
  \usepackage{graphicx}                
  \DeclareGraphicsExtensions{.eps}     
\fi%

\graphicspath{{figures/}{pictures/}{images/}{./}} 

\usepackage{microtype}                 
\PassOptionsToPackage{warn}{textcomp}  
\usepackage{textcomp}                  
\usepackage{mathptmx}                  
\usepackage{times}                     
\usepackage{cite}                      
\usepackage{tabu}                      
\usepackage{booktabs}                  

\onlineid{1801}

\vgtccategory{Research}

\vgtcinsertpkg

\preprinttext{Accepted at IEEE VR 2023}

\usepackage{dirtytalk}
\usepackage{hyperref}
\hypersetup{hidelinks}
\usepackage{booktabs}
\usepackage{hhline}
\usepackage{subcaption}
\usepackage{microtype}
\usepackage{multirow}
\usepackage{graphicx}
\usepackage[bottom]{footmisc}
\usepackage{todonotes}
\usepackage{breqn}
\usepackage{amsmath}
\usepackage{url}
\usepackage{comment}
\usepackage{ragged2e}
\usepackage{gensymb}

\newcommand\revision[1]{#1}
\newcommand\remove[1]{}

\title{MAGIC: Manipulating Avatars and Gestures to Improve Remote Collaboration}

\author{Catarina G. Fidalgo\thanks{e-mail: cfidalgo@andrew.cmu.edu}\\ %
        \parbox{1.4in}{\scriptsize \centering Carnegie Melon University \\ Instituto Superior Técnico, University of Lisbon}\\%
\and Maurício Sousa\thanks{e-mail: mauricio.sousa@dgp.toronto.edu}\\ %
     \scriptsize University of Toronto %
\and Daniel Mendes\thanks{e-mail: danielmendes@fe.up.pt}\\ %
     \parbox{1.4in}{\scriptsize \centering INESC TEC \\ Faculdade de Engenharia, University of Porto}\\%
 \and Rafael Kuffner dos Anjos\thanks{e-mail: r.kuffnerdosanjos@leeds.ac.uk}\\ %
     \scriptsize University of Leeds %
\and Daniel Medeiros\thanks{e-mail: daniel.piresdesamedeiros@glasgow.ac.uk}\\ %
     \scriptsize University of Glasgow %
\and Karan Singh\thanks{e-mail: karan@dgp.toronto.edu}\\ %
     \scriptsize University of Toronto %
\and Joaquim Jorge\thanks{e-mail: jorgej@tecnico.ulisboa.pt}\\ %
     \parbox{1.4in}{\scriptsize \centering INESC-ID \\ Instituto Superior Técnico, University of Lisbon}\\%
     }

\teaser{
  \centering
      \includegraphics[width=0.85\linewidth]{figures/1-MAGIC-teaser.pdf}
      \caption{Collaborating remotely in a shared 3D workspace: A) \textit{Veridical Face-to-face Remote Meeting}: Kate and George can see each other but have opposing points-of-view of the workspace. Kate indicates the optimal position for the component she is designing using proximal pointing gestures. George does not understand the location Kate is referring to as it is occluded. Kate needs to explain it through words or ask George to come by her side,  which brings added complexity to the collaboration task.
      B) \textit{MAGIC Remote Meeting}: Kate and George both share the "illusion" of standing on opposite sides of the workspace but are in fact sharing the same point-of-view. Kate's representation is manipulated so that George sees her arm pointing to the location she intends to convey. George understands where she is referring to, without any added effort.}
      \label{fig:teaser}    
}

\abstract{

Remote collaborative work has become pervasive in many settings, ranging from engineering to medical professions.
Users are immersed in virtual environments and communicate through life-sized avatars that enable face-to-face collaboration. 
Within this context, users often collaboratively view and interact with virtual 3D models, for example to assist in the design of new devices such as customized prosthetics, vehicles or buildings.
Discussing such shared 3D content face-to-face, however, has a variety of  challenges such as ambiguities, occlusions, and different viewpoints that all decrease mutual awareness, which in turn leads to decreased task performance and increased errors.
To address this challenge, we introduce MAGIC, a novel approach for understanding pointing gestures in a face-to-face shared 3D space, improving mutual understanding and awareness.
Our approach distorts the remote user’s gestures to correctly reflect them in the local user's reference space when face-to-face. 
To measure what two users perceive in common when using pointing gestures in a shared 3D space, we introduce a novel metric called pointing agreement.
Results from a user study suggest that MAGIC significantly improves pointing agreement in face-to-face collaboration settings, improving co-presence and awareness of interactions performed in the shared space.
We believe that MAGIC improves remote collaboration by enabling simpler communication mechanisms and better mutual awareness.

} 

\CCScatlist{
  \CCScatTwelve{Human-centered computing}{Collaborative and social computing}{Collaborative and social computing theory, concepts and paradigms}{Computer supported cooperative work}
}

\begin{document}



\maketitle

\section{Introduction}

In our globalized world, remote collaboration is increasingly important. 
Air travel cost and time overheads, global warming, and more recently, the pandemics made the ability to work remotely very popular, if not essential.

Traditional approaches to remote collaboration, using voice and video, have been used for decades.
However, these technologies neglect essential aspects of interpersonal nonverbal communication such as 
gaze, body posture, gestures to indicate objects referred to in speech (\textit{deictic gestures}~\cite{mcneill1992hand}), or even how people use the space and position themselves when communicating (i.e. \textit{proxemics}~\cite{hall1966hidden}). 
These limitations negatively affect how remote people interact with a local user, especially for operations that require 3D spatial referencing and action demonstration~\cite{10.1145/3089269.3089281}.


\textit{ Mixed Reality} (MR) is a key technology for enabling effective remote collaboration. Fields such as engineering, architecture, and medicine already take advantage of MR approaches for 3D models' analysis. Remote collaboration through MR allows teams to meet in life-sized representations and discuss topics as if sharing the same office. This enables a high level of co-presence and allows a more rigorous spatial comprehension of the 3D models they are working on \cite{krolovitsch20093d,brunet2015immersive}. Indeed, previous research found that for a remote meeting to be closer to a co-located experience, it should rely on a real size portrayal of remote people to maintain the sense of ``being there”~\cite{pejsa2016room2room, Orts-Escolano:2016:HVT:2984511.2984517,Gotsch:2018:TCL:3173574.3174096}. 
Furthermore, people perform tasks better when 
communicating via full-body gestures \cite{dodds2010communication}.

A \textit{face-to-face f-formation}, i.e., two people position themselves facing each other~\cite{fformations}, was found to have potential advantages over \textit{side-by-side f-formations}, i.e., two people position themselves side to side facing the same direction~\cite{fformations}, including gaze awareness, shared attention and the ability to see the other person’s visual cues ~\cite{Zillner:2014:WRC:2642918.2647393, Ishii:1992:CSM:142750.142977}. In addition to hindering the capacity to recognize consequential and intentional communicative cues, in a side-by-side f-formation participants share the same personal space, which makes it hard to implement mechanisms that can coordinate the use of this shared space \cite{Ishii:1992:CSM:142750.142977}. 
However, 
when virtually collaborating in a ``veridical face-to-face setting", people have to deal with the problem of sharing opposing points of view (POVs) of the workspace. In these situations, people do not share the same \textit{forward-backward} orientation, and there can be occlusions of parts of the workspace. 
This constrains the ability for people to use descriptions of relative positions and affects the understanding of where or what the remote person is pointing at, leading to communication missteps and causing tasks to be more laborious~\cite{negativeSpace, Zillner:2014:WRC:2642918.2647393}. 
This problem of disjoint visual perspectives when referring the workspace has been identified in previous work~\cite{whittaker2003things, Ishii:1992:CSM:142750.142977, Wood:2016:SHR:2992154.2992169, Li:2014:ITT:2598510.2598518,  Zillner:2014:WRC:2642918.2647393}, especially when using pointing gestures~\cite{tatar1991design, hindmarsh2000embodied}. 
Previous work found that for 3D-object-centered collaboration, a shared POV is more effective and preferred when compared to an opposing POV ~\cite{feick2018perspective, vishnu}.
\revision{Although shared POV is a well-established approach for this purpose,} 
\remove{However,} users lose the communicational advantages of seeing their partner in front of them~\cite{ishii1994interactive}\revision{, being deprived of non-verbal cues such as gaze, facial expressions, and body posture which are essential in terms of awareness of all remote users involved in the collaborative task. Indeed, Gutwin and Greenberg \cite{gutwin2002descriptive} suggest that to ensure workspace awareness people should gather information in ways familiar to what happens in real environments, with mechanisms such as {consequential communication}, {feedthrough}, and {intentional communication}. This first source of information, \textit{consequential communication}, relies on the other person’s body: posture, movement of head, eyes, arms, and hands, or facial expressions provide an abundance of information about what is happening in the workspace.}

We conclude that shared viewpoints and face-to-face dialogues are both beneficial but incompatible.
\revision{To address this challenge, we explore integrating both by manipulating user’s perception of each other and the environment, to provide a novel share of perspective where the local user is also aware of their partner’s consequential communication. }

\remove{In this work} \revision{Hence}, we present MAGIC, a novel approach to collaborative remote work that enables two people to interact using pointing gestures, adopting a face-to-face f-formation while sharing a common perspective of the workspace. 
MAGIC distorts gestures performed by a remote person to present a corrected virtual image to the local participant, ensuring accurate communication. 
Prior work suggests that the interpretation of pointing gestures can be significantly improved by warping the pointing arm~\cite{mayer2020improving,warpingdeixis}. Sousa et al.~\cite{warpingdeixis} developed a model for correcting vertical errors using targets within one fixed column (1D), while Mayer et al.~\cite{mayer2020improving} manipulated pointing gestures for two dimensions. However, these works are focused on distal pointing for side-by-side f-formations with targets at a fixed distance. 
Our work focuses on proximal pointing for face-to-face f-formations where targets can be anywhere in the shared 3D space. 
Collaborators can use pointing gestures to refer to proximal objects in the workspace and interpret their partner's movements in a modified reference space. This creates the illusion that they performed a gesture as expected since it corresponds to the reference space of the local participant, as demonstrated by the usage scenario in Figure~\ref{fig:teaser}. 
We implemented a telepresence environment using virtual representations of people rendered life-size around a shared workspace to evaluate our approach. Additionally, we introduce \textit{Pointing Agreement} to classify what participants perceive in common when using proximal pointing gestures to indicate targets located in a shared workspace, measured through the application of the \textit{Jaccard Similarity Coefficient}~\cite{jaccard1912distribution, tanimoto1968elementary} in a 3D space. 
The user study results showed improved capabilities to understand pointing gestures using MAGIC as compared to a veridical face-to-face f-formation. Furthermore, participants did not detect our manipulations, and MAGIC distortions did not negatively affect the participants' sense of presence or the time they spent doing a task.
Our experiments validated the assumption that sharing the same perspective while in a face-to-face f-formation improves \textit{pointing agreement} and facilitates virtual object-centered remote collaboration.

The main contributions of this research include: 
1) MAGIC, a novel interaction technique to integrate person-, task-, and reference spaces to improve the understanding of proximal pointing gestures in face-to-face remote collaboration in virtual environments; 
{
2) Pointing Agreement, a concept that encapsulates referential consensus between two face-to-face collaborators;} 
3) A user study, demonstrating the effectiveness of our technique.


\section{Related Work}

MAGIC builds on research in workspace awareness in shared spaces, interpretation of pointing gestures in Collaborative Virtual Environments (CVEs), and perception manipulation techniques.

\subsection{Workspace Awareness in Shared Spaces}
Workspace awareness relates to the immediate changes that occur in a shared workspace, and can be defined as the \textit{"up-to-the-moment understanding of another person's interaction with a shared workspace"}~\cite{gutwin2002descriptive}.
Greenberg et al.~\cite{greenberg1996awareness} suggest that 
this real-time knowledge of another person's interactions and their effects on the workspace is essential to 
effective collaboration. 
As working together causes people to undertake the additional task of maintaining 
collaboration through communication and decision making, apart from their individual domain jobs~\cite{adams1995situation}, 
inadequate workspace awareness 
causes people to perform more challenging and awkward collaboration tasks, which, in turn, causes the domain tasks to be more laborious~\cite{negativeSpace}.

In addition to mechanisms such as \textit{feedthrough}, \textit{consequential communication} and \textit{intentional communication} 
being challenging to recreate when collaborating remotely,  
 most telepresence systems typically rely on a separation between the \emph{person space} and the \emph{task space}. 

Buxton \cite{buxton1992telepresence} suggested that \textit{"effective telepresence depends on quality sharing of both person and task space"}, with collaborators meeting in a shared space where they can refer 
the workspace using gaze or deictic gestures, the \emph{reference space}.

An early example of this concept is the Clearboard ~\cite{Ishii:1992:CSM:142750.142977}, where two participants engage in collaborative drawing tasks while seeing each other face-to-face. 
To correct for the inaccurate reference system, the authors resorted to horizontally reverse the video streams to establish the same point-of-view for both participants. 
To enable users to collaborate in a common reference space, other approaches combined physical and virtual objects on arbitrary surfaces~\cite{junuzovic2012see}, rendered the remote user's hands on the local space~\cite{sodhi2013bethere}, or mirrored remote users' representations rendered on top of a whiteboard content~\cite{Zillner:2014:WRC:2642918.2647393}. Leithinger et al. enabled users to manipulate physical shapes, with the person space in a face-to-face or corner-to-corner f-formation on the sides of the shared workspace~\cite{Leithinger:2014:PTS:2642918.2647377}. These demonstrated that face-to-face telepresence approaches allow for improved effectiveness when compared to side-by-side settings~\cite{Zillner:2014:WRC:2642918.2647393}, and suggested that to maintain awareness in face-to-face interactions, telepresence systems should resort to selective image reversal of text and graphics~\cite{Li:2014:ITT:2598510.2598518}.

\revision{Speicher et al.~\cite{speicher2018} provided users with a third-person view by default but enabled one collaborator at a time to gain control over everyone’s 360\degree  video. This enabled all collaborators to view directions in a synchronized manner. 
Similarly, Teo et al. \cite{teo2019} allowed experts to interact with trainees in a third-person view by default, but with the option to immerse themselves into the same point of view as the local collaborator, live-streamed through a 360\degree camera attached to the trainee’s head.
Cai et al. \cite{cai2017} demonstrated a system that shows the remote user’s virtual head and hands in a shared live 360 panorama captured from a backpack-mounted 360 camera. 
Lee et al. \cite{lee2017} created a system that allows sharing of hand gestures and awareness cues over live 360 panoramas captured by a head-worn camera, enabling two-way non-verbal communication between a pair of users wearing AR or VR displays. 
Similarly, Teo et al. \cite{teo2018} investigated adding ray pointing and drawing annotations to hand gestures as non-verbal communication cues in a live 360 panorama-based MR remote collaboration system, showing that participants acting as a local users were able to perform tasks faster and with less error with the help of visual annotation cues.
On the other hand, Rhee et al.~\cite{rhee2020augmented} created a system for enabling face-to-face collaboration using 360 panoramas that enabled a remote user to be teleported to a visually rich remote location with a visually-matched virtual object and share their interaction intentions with pointing gestures and voice.}

Regarding 3D object manipulation, Benko et al.~\cite{Benko:2012:MFI:2207676.2207704} introduced MirageTable, 
to bring people together face-to-face as if they were working 
at the same table.
Participants share the same task space, interacting with physical objects, although this approach did not seemingly avoid occlusions. 
Additionally, \revision{Kim et al. \cite{kim2019} explored the combination of different visual cues (e.g., pointer, sketching, and hand gestures) showing a significantly higher level of usability when the sketch cue is added to a hand-pointing gesture cue alone.}

We conclude that face-to-face encounters provide benefits to remote collaboration, although pointing at 3D objects from different perspectives still induces ambiguities since collaborator's gestures may be occluded. We tackle this challenge by implementing task- and personal spaces in combination. We also hypothesize that enabling the same perspective while correcting people's actions can improve intentional and consequential communication mechanisms mechanisms as it can make users' gestures and other visual actions more understandable by both parts of a remote encounter.
\vspace{-0.1 cm}

\subsection{Manipulating People's Perception}

In virtual meetings, the sense of presence of remote users has an essential role in the meeting participant's capacity to communicate and collaborate. Previous research suggests that utilizing complete or upper-body representations improves people's awareness~\cite{Benford:1995:UEC:223904.223935, buxton1992telepresence} since a richer vocabulary combining body language with speech can be used. Furthermore, understanding the other person's gaze~\cite{nguyen2005multiview}, communicative gestures~\cite{Bekker:1995:AGF:225434.225452, kirk2005ways} and deictics~\cite{Genest2011, tang2007videoarms} are known to improve remote collaboration. 
To improve how one perceives the remote participants' intentions and actions, different warping techniques has been used by taking advantage of manipulating human perception.

A classic example of this approach is the Redirected Walking~\cite{razzaque2005redirected}, where the virtual environment is transformed so that the person can use a natural metaphor to cover larger distances in a limited physical space~\cite{azmandian2016haptic}. 
Besides the environment, retargeting techniques can be applied to the representation of people.
This was first introduced with inverse kinematics to animate virtual characters when incomplete motion was available~\cite{gleicher1998retargetting,lee1999hierarchical}, and for one single person to simultaneously animate multiple skeletons~\cite{hecker2008real}. Feuchtner et al.~\cite{feuchtner} extended the users' arm to allow for access to devices that are out of reach.
Other approaches use motion remapping to modify real people's poses to match the motion of a person in a given video~\cite{chan2018everybody}. 
Recent work also suggests that the interpretation of deictic gestures can be significantly improved by using retargeting techniques that warp the pointing arm~\cite{Zillner:2014:WRC:2642918.2647393,mayer2020improving,warpingdeixis}.


Piumsomboon et al. \cite{miniMe} found that manipulating the remote user's representation in order to guarantee that his gestures were always in the field of view of the local user enables a high sense of presence and increases task performance over an unmodified avatar. 
Sousa et al. \cite{negativeSpace} studied different manipulations of remote people and workspaces, mirroring one or the other in order to enable a shared point of view for the users. Results suggested remote collaboration benefits more from workspace consistency rather than people’s representation fidelity.
Hoppe et al. \cite{hopeetal} also manipulated virtual people's representation to provide a shared point of view, finding that modifications of the workspace and the user’s avatar to induce a shared perspective reduces mental load and increases task performance. And later, in \cite{shisha}, Hope et al. further explore shared perspective by comparing the baseline condition of two users standing in the same location and working with overlapping avatars, with their approach of shifting the remote collaborator's representation to the side and redirecting their gestures to match the reference space using laser pointing.

In MAGIC, we warp the remote collaborator's representation to guarantee that their gestures match the reference space when enabling a share of perspective. Similarly to~\cite{gleicher1998retargetting,lee1999hierarchical,miniMe}, we apply inverse kinematics when redirecting the remote avatar's gestures.\\

\vspace{-0.5 cm}

\subsection{{Interpreting Pointing Gestures in CVEs}}
\label{rw_pointing} 
The problem of interpreting remote pointing gestures in collaborative tasks has been long considered.
In the context of 2D gestures, Greenberg et al.~\cite{ greenberg1996semantic} considered the problem of representational differences between two collaborators’ views when using 2D telepointers, outlining potential solutions such as scaling or object-based pointing. 
Hindmarsh et al.~\cite{hindmarsh1998fragmented,hindmarsh2000object} also focused on the difficulties in identifying what other people are pointing at in a VR environments, identifying difficulties resolving pointing gestures when the pointing embodiments and target objects were not both in view. As a possible solution, they suggested to focus on exaggerating the representation of actions so that they can easily been seen by others. 
Wong et al. extended Hindmarsh et al's work to quantify the amount of error in people’s interpretation of others’ gestures~\cite{wong2010you, wong2014support} when pointing in distance. Similarly, Sousa et al. \cite{warpingdeixis}, and later Mayer et al. \cite{mayer2020improving}, proposed solutions for correcting this problem by using retargeting techniques. However, these works are focused on \textit{distal pointing} for \textit{side-by-side f-formations} with targets at a fixed distance along 1D \cite{warpingdeixis} or 2D \cite{mayer2020improving}. In our work, we address the problem of \textit{proximal occluded pointing in face-to-face scenarios} for targets anywhere in the 3D space reachable by the user.

\subsection{Discussion}

We found approaches~\cite{Zillner:2014:WRC:2642918.2647393, higuchi2015immerseboard,Ishii:1992:CSM:142750.142977,Li:2014:ITT:2598510.2598518} successful in ensuring all three types of workspace awareness mechanisms for 2D artifacts, and approaches~ \cite{Benko:2012:MFI:2207676.2207704,Leithinger:2014:PTS:2642918.2647377,beck2013immersive} that support 3D artifacts, but have a limited capacity to ensure intentional communicational cues, hindering the use of pointing gestures to refer the workspace.
To the best of our knowledge there is no approach for remote face-to-face object-centered collaboration with 3D content that allows all three types of workspace awareness mechanisms.

Analysing Ishii et al.~\cite{Ishii:1992:CSM:142750.142977}'s view on the problem of how to communicate in a shared environment without separation between the task and the personal space, we 
concluded that the \emph{over the table metaphor} would be the most suitable for object-centered collaboration. We believe that by enabling a share of perspective between collaborators, we can tackle the challenge of mismatched views of the task space and occlusions present in approaches with a similar metaphor~\cite{Benko:2012:MFI:2207676.2207704,Leithinger:2014:PTS:2642918.2647377, beck2013immersive}. 
We resort to manipulations of the remote person’s representation, as supported by \cite{negativeSpace,hopeetal}, to enable a
face-to-face meeting paired with a shared perspective of the workspace.
Regarding the interpretation of pointing gestures in CVEs, we build on top of previous work~\cite{wong2010you, wong2014support, warpingdeixis, mayer2020improving} exploring 
retargeting techniques to correct distal pointing interpretation in 1D and 2D, to address the problem of proximal occluded pointing in 3D face-to-face scenarios. Similarly to~\cite{warpingdeixis,gleicher1998retargetting,lee1999hierarchical,miniMe, mayer2020improving}, we apply inverse kinematics when redirecting the remote avatar's gestures.

\begin{figure*}[t!]
\centering
\includegraphics[width=\textwidth]{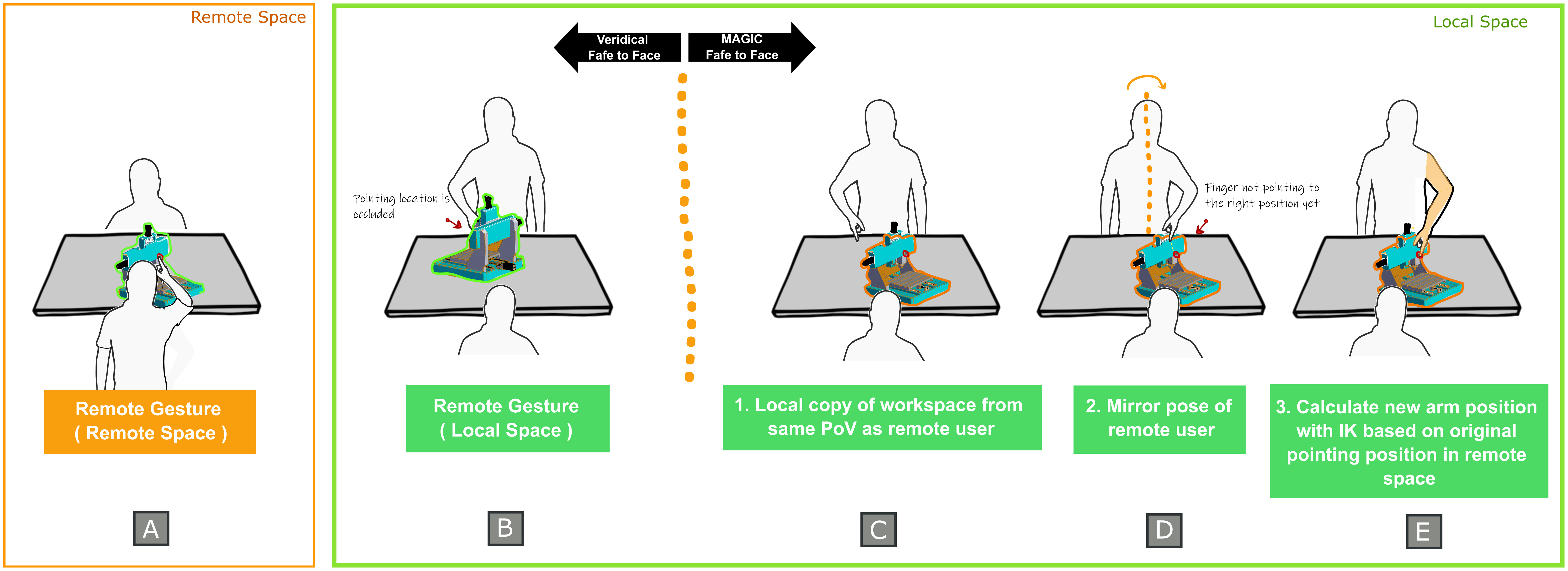}
\caption{Illustration of the different steps involved in implementing our approach.}
\label{fig:steps}
\end{figure*}

\section{MAGIC}


Following the assumptions of Gutwin and Greenberg~\cite{gutwin2002descriptive} that peripheral tasks bring additional efforts in maintaining collaboration, we propose to diminish the shift of attention between the personal space and the task space by creating a virtual environment to integrate the two, improving workspace awareness. To achieve that, MAGIC employs body representation manipulations in ways unnoticed by users to improve the understanding of pointing gestures used in the shared space, letting users focus on the main collaborative task. 
The present work addresses collaboration with two users only. While this is arguably the simplest case, future work could explore the extension of our work's principles to accommodate multiple people. In the following, we detail the design and implementation of MAGIC.

\subsection{Improving Workspace Awareness}
\label{ImprovingWA}

MAGIC assumes an "above-the-table" metaphor~\cite{Ishii:1992:CSM:142750.142977} with life-sized virtual representations of remote people in front of one another with the workspace between them, as depicted in Figure~\ref{fig:teaser}.
In a face-to-face f-formation, participants can perceive natural gestures besides maintaining verbal communication, which contributes to 
presence and increases awareness of other people's actions. This reduces the separation between task space and personal space. Hence, it enables the use of pointing gestures when interacting with the workspace's virtual artefacts. Yet, the 3D virtual model's occlusions can diminish workspace awareness. To tackle this problem, MAGIC guarantees that the two participants share the same point of view of the workspace at all times.
Sharing the same perspective in object-centred remote collaboration allows users to have a mutual understanding of the task space, allowing for the use of natural communication, easing the pressure of engaging in lengthy verbal cues about task-specific points of interest~\cite{feick2018perspective}.

However, to allow for a face-to-face setting with the same perspective in shared 3D workspaces, we can't only render the remote person in front of the local collaborator.
This naive approach makes the remote collaborator's gestures spatially incorrect as their gestures will not match the reference space.
To maintain the same reference space, MAGIC manipulates the remote person's representation so that there is a matching between the local observer's pointing location and the point of interest to which the remote person is referring. 
Therefore, MAGIC aims to increase workspace awareness in object-centered three-dimensional collaboration by improving the pointing agreement between two collaborators. 
It integrates person-, task- and reference spaces, enabling collaborators to thoroughly express and perceive consequential communication, feed-through, and intentional communication.

\noindent\textbf{Consequential Communication:}
Since MAGIC adopts a life-sized face-to-face arrangement, the remote collaborator is always visible, similarly to traditional face-to-face interactions. 
Therefore, the local person can observe posture and body language, inferring significant information about what is happening in the workspace. 
This way, remote collaborators do not need to switch 
back-and-forth between the person space and the task space (requiring fewer eye-movements and no body rotations away from the workspace).

\noindent\textbf{Feedthrough:}
When manipulating artifacts in the workspace, they provide feedback information to both the person performing the action and the observer.
Thus, sharing the same perspective of the workspace allows direct object manipulations to 
retain position and orientation in both the local and remote workspaces, assuring that feedthrough.
    
\noindent\textbf{Intentional Communication:}
MAGIC assures that gestures in the remote workspace are correctly converted to the local reference space.
And since both participants have the same understanding of the workspace artifacts' whereabouts, pointing gestures performed remotely remain meaningful when converted, allowing collaborators to communicate using natural language and gesture freely and accurately.

\vspace{-0.2 cm}

\subsection{Manipulating users' virtual representation}
To combine the advantages of both being face-to-face and sharing perspective, both participants stand on the same location, sharing the same point of view of the 3D model, but see a manipulated version of their partner. 
These manipulations include 
mirroring along 
left-right 
(z-axis), similarly to Zillner et al.~\cite{Zillner:2014:WRC:2642918.2647393} followed by an adjustment of the remote person's overall pose to correct for the depth of the interaction.

In Figure~\ref{fig:steps}, we illustrate the different steps necessary to accomplish the manipulations that enable our approach, where a remote collaborator intentionally communicates the position of a workspace artifact using a proximal pointing gesture.

In their local space, the remote collaborator points to an area of interest within the shared workspace to communicate a specific area of the 3D artifact (Figure \ref{fig:steps} [A]). 
In a veridical face-to-face scenario, the remote pointing gesture is occluded from the local participant (Figure \ref{fig:steps} [B]).
With MAGIC, we follow the following steps to make the remote pointing gesture visible by the local user.

\noindent\textbf{1. Workspace Local Copy:}
We first procede to the renderization of the workspace artifact from the same PoV as the remote user their local space. In the local space, without any manipulation, the remote user's gesture highlights a different area of the task space (Figure \ref{fig:steps} [C]).

\noindent\textbf{2. Mirror:}
Similarly to Clearboard~\cite{Ishii:1992:CSM:142750.142977} and 3D-Board~\cite{Zillner:2014:WRC:2642918.2647393}, we then 
mirror 
the remote person’s representation (Figure \ref{fig:steps} [D]). 
A local observer can correctly understand the remote person's interactions in the workspace's left-right axis after mirroring. 
There is a common understanding of right, left. Body language matches horizontally 
in the local reference space.  

\begin{figure}[b!]
\centering
\includegraphics[width=\columnwidth]{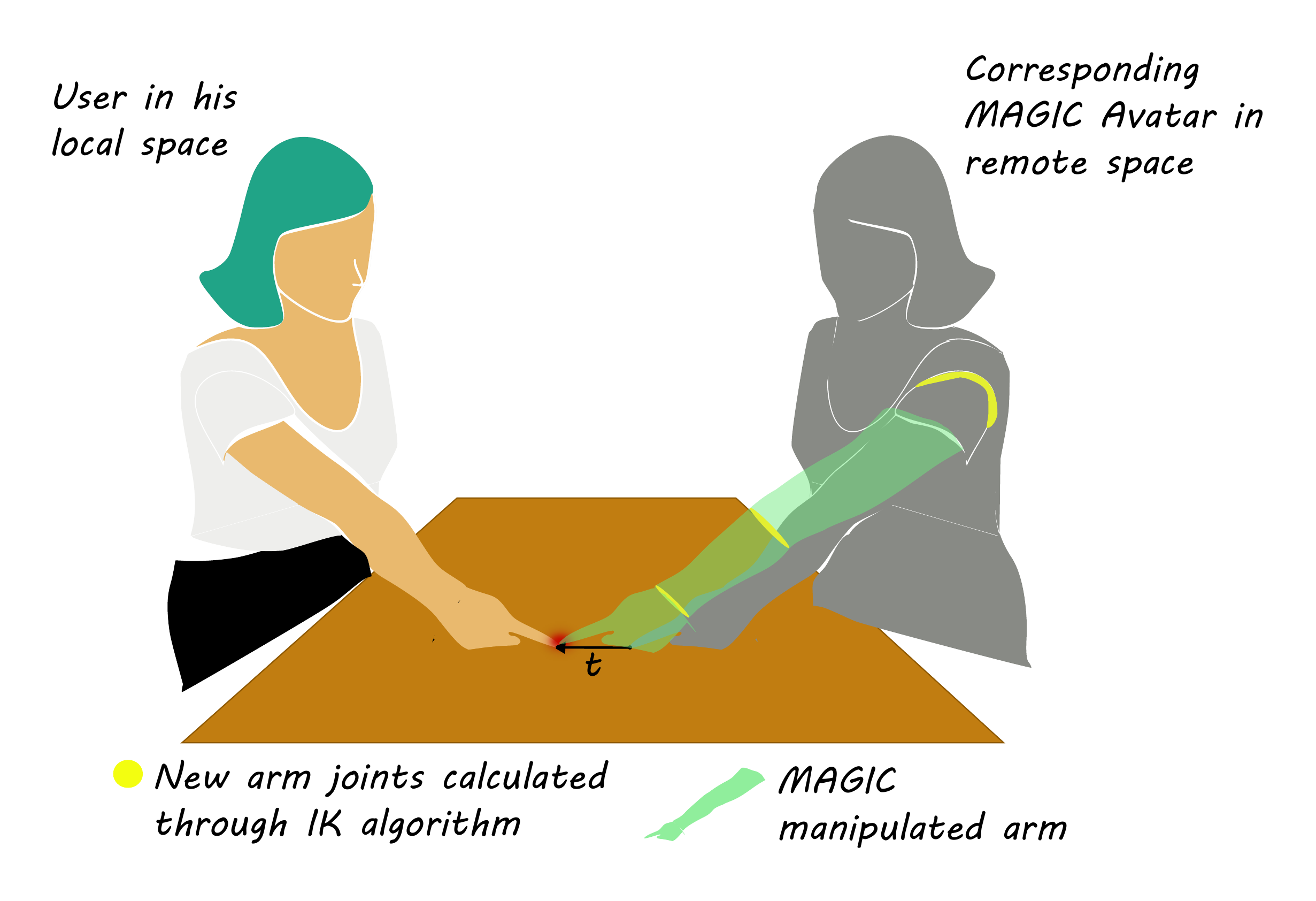}
\caption{Manipulated avatar (right) corresponding to a pointing user in his local space (left). Transformation $\vec t$ is calculated through the difference in position from the remote collaborator’s fingertips in his local space (left), and the remote avatar’s fingertips after the mirroring process (right - grey). $\vec t$ is applied to the remote avatar’s fingertip in combination with IK algorithm to calculate new arm joint's positions. }
\label{fig:poitingarm}
\end{figure}

\noindent\textbf{3. Adjust Arm Pose:}
Since we are working in a 3D workspace, the mirroring by itself is insufficient, and depth needs to be corrected.
We re-position the pointing arm along the forward-backward axis to correct for the depth of the interaction (Figure \ref{fig:steps} [E]). The transformation vector, $\vec t$, applied to the remote avatar’s pointing finger, is calculated through the difference in position from the remote collaborator’s fingertips in his local space, and the remote avatar’s fingertips after the mirroring process (Figure \ref{fig:poitingarm}).
We note that all the arm segments need to move when changing the pointing finger's end position. We need to displace the upper arm, the forearm, and the wrist according to the finger's new position. 
We can calculate these movements using Inverse Kinematics constrained by shoulder, elbow and wrist joints. As we focus on proximal pointing, the finger direction itself does not influence understanding of position as long as the fingertip is in the right place~\cite{UpToTheFingertip}.

Additionally, if the remote person is in a position relative to the worktable where, in order to match the workspace's pointing position, requires stretching of the arm beyond the arm length, we add an additional step: moving the remote person's avatar further along the forward-backward axis. 
An arm that is too long could make the avatar representation weird for the local observer, breaking the user's sense of presence~\cite{mori2017uncanny}.
As we want to adjust the avatar’s position, $p_a$, according to the remote collaborator’s index fingertip position, $p_f$ , we compute the remote avatar's head position as a linear function of the updated fingertip position: 
\begin{equation}
\label{eq:1}
    p_a = m \cdot p_f + b,
\end{equation} 
where m and b are calculated based on the limit points of the shared space considered (the table dimensions in our case).

\section{Evaluation}




We conducted a user study to determine whether our approach improves face-to-face collaboration in 3D object-centered remote collaboration settings 
when both participants share the same perspective, when compared to a \textit{veridical} face-to-face.
Therefore, our hypothesis throughout the experiment is:


\textit{
\label{hyp:one}
\textbf{$H$}: A face-to-face setting coupled with manipulations that enable perspective sharing improves the understanding of proximal pointing gestures, increasing workspace awareness.}

Additionally, as we designed our approach to be used implicitly, we also wanted to confirm that it did not distract users from 
the main 
task. Furthermore, we wanted to assess whether the shared point of view with corrected gestures impacts the feeling of presence of the remote collaborator.

To test our hypothesis, we conducted a within-subjects user study where we asked pairs of participants 
to complete a collaborative pointing task under two different conditions:

    \noindent \textbf{1. MAGIC Face-to-Face}: Participants can see their partner located in front of them, both share the "illusion" of standing on opposite sides of the workspace but are in fact sharing the same point of view. Remote participant's representation is manipulated in order to enable the illusion.
    
    \noindent \textbf{2. Veridical Face-to-Face}: Participants can see their partner located in front of them, each one standing on opposite sides of the model, with opposing points of view of the workspace, as it occurs in "traditional" real life face-to-face interactions. 

These two conditions were designed to allow us to compare MAGIC interactions with "traditional" face-to-face collaboration, with the goal of improving face-to-face collaboration specifically. We did not compare or address mirrored side-by-side settings, since these prevent visualizing both the workspace and the opposite participant, making it difficult to take advantage of gaze direction, posture and body language cues. Additionally, in order to use proximal pointing, side-by-side collaborators need to invade each other’s personal space which is avoided if standing face to face as the table delimits each other’s personal spaces. Otherwise, collaborators need to stick to distal pointing which has been shown \cite{warpingdeixis} to induce errors in MR if uncorrected. Hence, we discarded side-by-side configurations and did not consider it in our experiments.

\subsection{Task}
\label{sec:eval_task}

\begin{figure}[t!]
    \centering
    \includegraphics[width=.95\columnwidth]{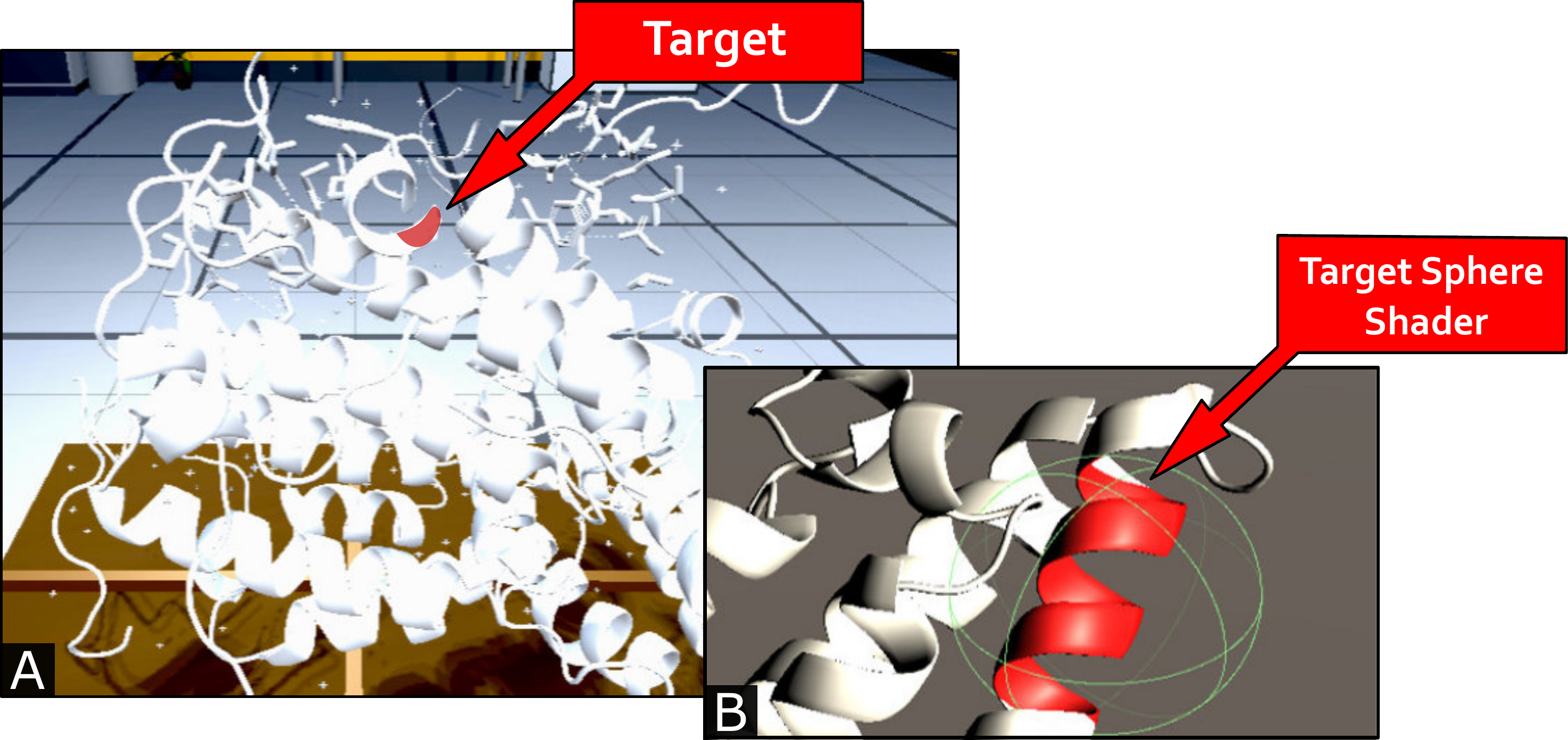}
    \caption{A) Abstract representation of a protein used for the task with red highlighted target to be indicated by the participants. B) Target sphere target.}
    \label{fig:task_model}
\end{figure}

We paired participants and asked them to complete a series of outlining tasks using proximal pointing. Each participant was assigned a different role in the beginning of the experiment, either \textit{Demonstrator} or \textit{Interpreter}.
The \textit{Demonstrator}'s role was to communicate a highlighted area of the task model to the \textit{Interpreter} by outlining it with his pointing finger. 
The \textit{Interpreter}, who could not see the highlighted area, was required to follow the \textit{Demonstrator}'s gestures and perform an outline of the interpreted area.

For the 3D model containing the targets participants had to indicate, we adopted an abstract representation of a protein, which was placed above the shared table (Figure \ref{fig:task_model}), visible to both participants. We employed an abstract 3D model so that it would be hard for participants to describe targets using only words and would be compelled to using pointing gestures. Participants could talk with each other freely throughout the experiment.


\subsection{Procedure}

All evaluation sessions followed the same structure, and each lasted for about 50 minutes.
Each session gathered two participants and the experiment moderator.
In response to the global circumstances caused by the COVID-19 outbreak, we undertook extra safety measures. 
Afterwards, participants completed a demographic profile questionnaire and a consent form. 
We then introduced the evaluation, where we explained conditions, tasks, and roles. 
Before executing the main task, participants had a training period to familiarize themselves with the environment and the hardware, which consisted of three outlining tasks. 

To start each matching task, the Demonstrator was asked to press a button on the controller, and once the workspace appeared, the target to communicate was shown as a small part of the workspace highlighted in red {(Figure \ref{fig:task_model} A)}.
The Demonstrator had to communicate that red target to their partner. To do so, the Demonstrator was asked to outline it with his finger while pressing a button from the controller, which would leave a green trail in the area outlined {(Figure \ref{fig:task_areas} A)}. This was only visible to the Demonstrator. The outline stopped once the button was released, ending the Demonstrator’s turn. 
Then, the Interpreter proceeded the same way, pressing a button while outlining his interpretation of the volume indicated by the Demonstrator, and finishing his turn by releasing the button.
Once an iteration finished, the target shown to the Demonstrator changed and the procedure was repeated. The Demonstrator was shown a set of 16 targets to communicate to their partner, one a time. Upon completion, we asked participants to answer a user preference questionnaire regarding the condition experienced.
Then, participants switched roles and performed the same task using a different set of 16 targets. 

Hence, each pair of participants performed under both conditions, and each participant experimented both roles (2 Conditions x 2 Roles x 16 Targets). As participants performed a similar task in more than one condition, we created four different sets of 16 targets and counterbalanced them to avoid carryover effects. Each target group comprised similar spots of the model that differed enough to guarantee that participants would not remember them from the precedent task but were similar enough to ensure consistency of tasks between participants and conditions. The order of conditions was also balanced.



\subsection{Evaluation Prototype}

\begin{figure}[t!]
    \centering
    \includegraphics[width=.95\columnwidth]{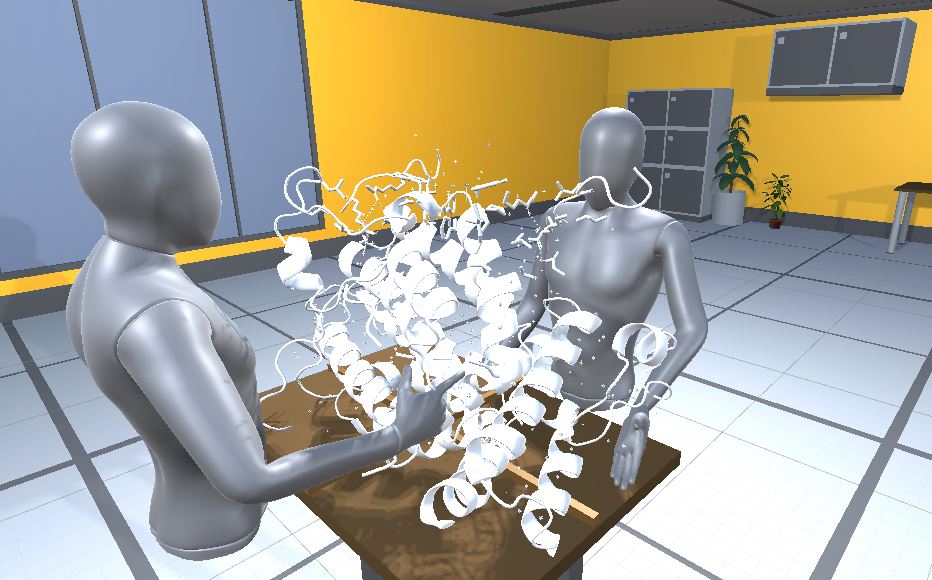}
    \caption{Our Prototype Virtual environment, 
    featuring two collaborators and the 
    abstract model used for the task.}
    \label{fig:prototype}
\end{figure}

We implemented our prototype using Unity3D. Although we envision our approach as relevant for both AR and VR, we conducted our evaluation in VR to minimize external factors and avoid problems resulting from matching real and virtual content in current AR hardware, and have a fully controllable experience. We created a simplistic office space with a tabletop and a 3D model between both users (see Figure~\ref{fig:prototype}). The prototype features an abstract avatar to embody each collaborator. We chose a humanoid abstract representation to avoid gender- and body-bias while focusing on nonverbal communication cues.
We used an Oculus Rift CV1 as the visualization tool, and we used 
the controllers' positions and rotations input to animate the avatars using inverse kinematics~\cite{Aristidou:2011:FABRIK,Aristidou:2016:ExtFABRIK}.
These data are then transmitted to the remote participant's prototype to animate the local avatar and display it on the remote participant's HMD. 
Whenever participants placed their hands inside the workspace, the virtual hand automatically adopted a pointing pose.

\begin{figure}[!b]
\centering
\includegraphics[width=.95\columnwidth]{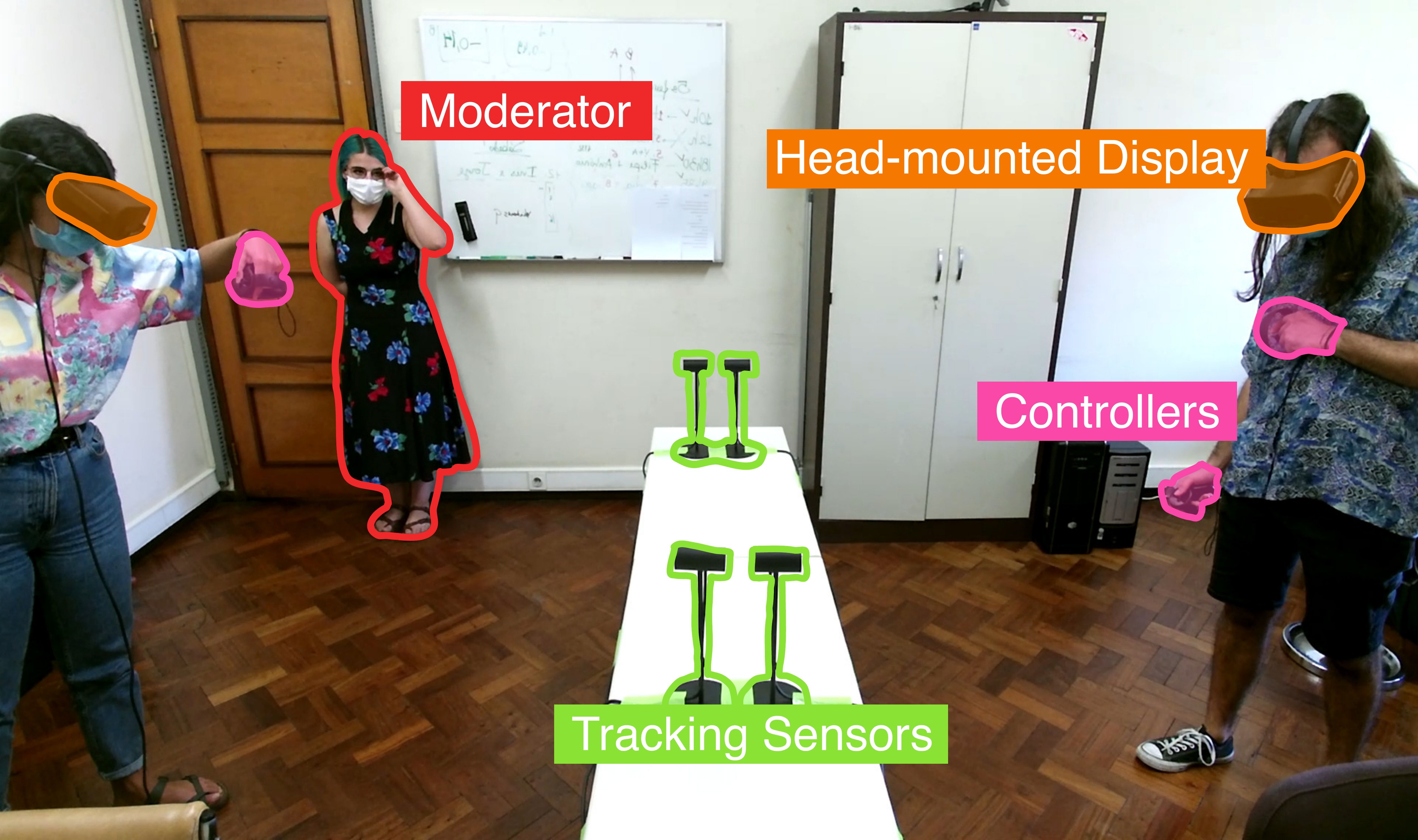}
\caption{Setup for the experiment: two users in different tracked stations wearing an Oculus Rift HMD and using Touch Controllers. Moderator was present at all times.}
\label{fig:setup1}
\end{figure}

\subsection{Setup and Apparatus}
Our experiment 
used a controlled laboratory environment with no contact with the exterior, occupied by the two participants and the 
moderator. The physical setup consisted of two stations 
separated by a physical divider. As participants were physically co-located, they could communicate verbally 
among themselves. Each station 
comprised a desktop running the participant’s application, and an Oculus Rift set (Oculus HMD, two Touch Controllers, and two-position trackers), as displayed in Figure~\ref{fig:setup1}.


\subsection{Participants}
We recruited 12 participants (3 female, 9 male) through convenience sampling in our institution. Participants were not compensated monetarily for participating in the user study. 
Participants’ ages ranged from 22 to 26 years (M = 23,6; SD = 1). 
All participants 
knew their partners. One participant reported being left-handed.
Three participants indicated previous experience with using video-conferencing platforms at least once a day, eight at least once a week, and one reported a monthly use. 
Three participants reported never having experienced VR environments, while the other nine participants reported rarely using 
such environments.

\section{Results}

Results from our user study suggest that MAGIC significantly improves \textit{pointing agreement} in face-to-face collaboration settings without negatively affecting the time to perform a task. Furthermore, it improves co-presence and awareness of interactions performed in the shared space. \\

\subsection{Collected Data and Metrics}
\label{metrics}

During the user evaluations we collected information on both Task Performance and User Preferences.

\noindent \textsc{\textbf{Task Performance}}
To evaluate task performance, we measured the task error by determining the \textit{pointing agreement} between both participants. Additionally, we measured the task completion time. 

\subsubsection{Pointing Agreement}
We define \textit{Pointing Agreement} as a measure of what two users perceive in common when using pointing gestures in a shared 3D space. We evaluate the pointing agreement between two users by calculating the \textit{Jaccard Similarity Coefficient}~\cite{jaccard1912distribution}. The Jaccard Similarity Coefficient, \textit{J}, is a metric usually used in understanding the similarities between sample sets, and is formally defined as the size of the intersection divided by the size of the union of the sample sets, as follows:

    \begin{equation}
        \textit{J}(A,B) = \frac{|A \cap B|}{|A \cup B|} = \frac{|A \cap B|}{|A| + |B| - |A \cap B|}.
        \label{eq:pointingagreement}
    \end{equation}

\begin{figure}[t!]
    \centering
    \includegraphics[width=.95\columnwidth]{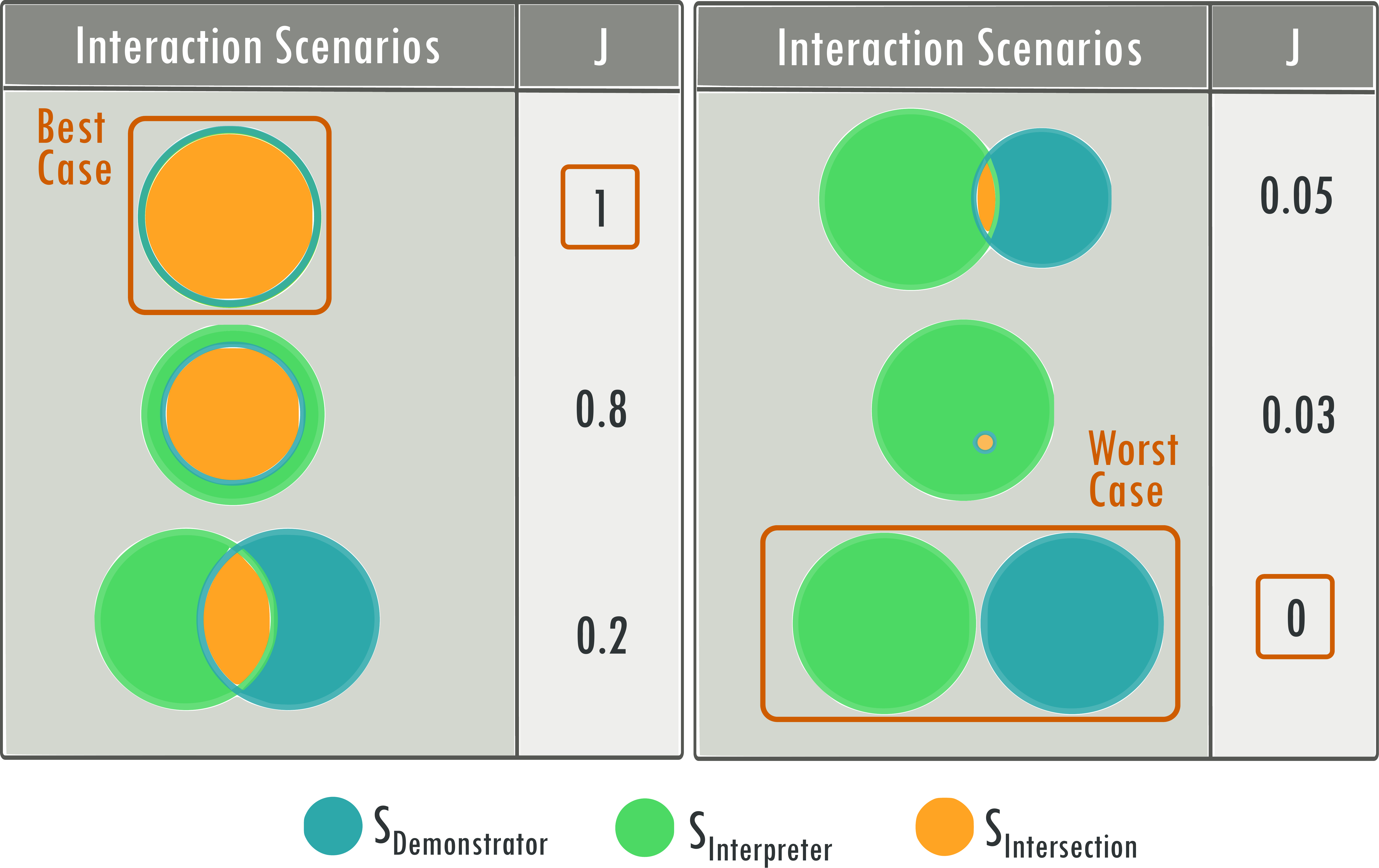}
    \caption{Different possible scenarios for outlines from best to worst case scenarios, and correspondent \textit{J}. Blue, green and orange represent the Demonstrator's Solid, $S_{Dem}$, Interpreter's Solid, $S_{Int}$, and the Intersection 
    of both, $S_{I}$ }
    \label{fig:PA_Scenarios}
\end{figure}

We apply this to the 3D sample sets of points that constitute the zone of interest pointed by each person. We chose this metric since it is able to characterize all possible scenarios in our experiments, as shown in Figure \ref{fig:PA_Scenarios}.
Hence, the higher the Pointing Agreement, the less error was present in identifying the region indicated by the task demonstrator and vice versa.
We note that, as we chose to designate parts of an abstract object, a centroid-based approach would not be accurate enough to show agreement as the target shapes are not spherical. Merely comparing 3D points would only measure the error distance and not highlight what participants had commonly perceived. 

\begin{figure}[!b]
    \centering
    \includegraphics[width=.95\columnwidth]{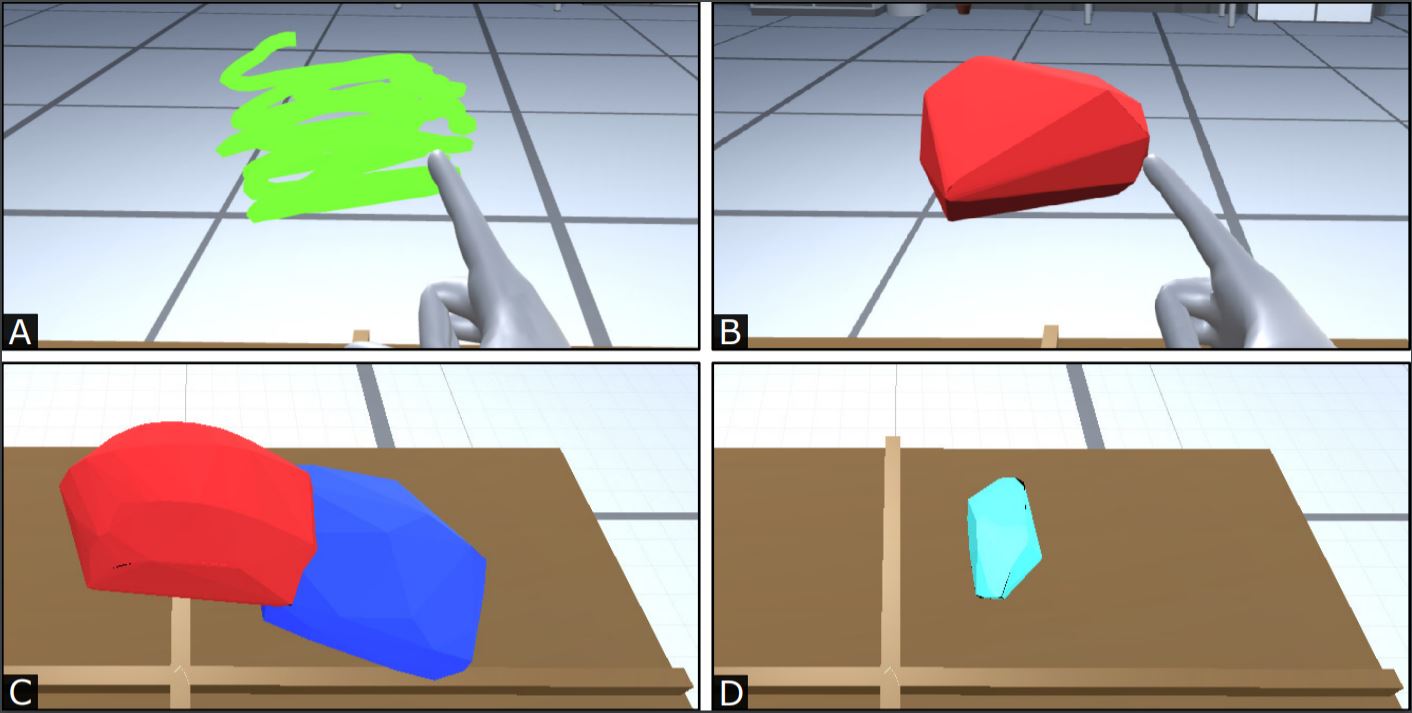}
    \caption{Calculating the intersection of the volumes outlined by both participants: (A) Outline of zone of interest; (B) Corresponding Convex Hull; (C) $S_{Dem}$ (red) and $S_{Int}$ (blue); (D) Resulting intersection, $S_{I}$. 
    }
    \label{fig:task_areas}
\end{figure}

To measure \textit{J} for each outlining task, we 
generate, for each participant,
the 
convex set comprising all the points that constitute the
outline of the zone of interest (Figure~\ref{fig:task_areas} A). This enables us to calculate the volume outlined by the Demonstrator, $S_{Dem}$ (Figure~\ref{fig:task_areas} C), the one the Interpreter perceived, $S_{Int}$ (Figure~\ref{fig:task_areas} C), and the volume of intersection between both, $S_{I}$ (Figure~\ref{fig:task_areas} D).

We use an adapted QuickHull Algorithm \footnote{Quickhull Algorithm for Generating 3D Convex Hulls - an Implementation in Unity, Oskar Sigvardsson (Accessed: 2020-05-15): \url{https://github.com/OskarSigvardsson/unity-quickhull}} to create $S_{Dem}$ and $S_{Int}$, and a \textit{Constructive Solid Geometry (CSG) Library}
\footnote{Constructive Solid Geometry (CSG) for Unity in C\#, Andrew Perry (2020-07-22): \url{https://github.com/omgwtfgames/csg.cs/blob/master/Assets/CSG/Plugins/CSG/CSG.cs}} to retrieve the intersection between the two, $S_{I}$.
We calculate the volume of the meshes, $V_{S_{Int}}$, $V_{S_{Dem}}$, and $V_{S_{I}}$, by adding the volumes of all tetrahedra that compose the mesh. The tetrahedra joins each triangle in the mesh with the vertex at the origin~\cite{TetrahedronVolume}. 
From expression \ref{eq:pointingagreement}, 
\textit{J} 
becomes:
    \begin{dmath}
        \textit{J}(S_{Int},S_{Dem}) = \frac{|S_{Int} \cap S_{Dem}|}{|S_{Int}| + |S_{Dem}| - |S_{Int} \cap S_{Dem}|} 
        = \frac{V_{S_{I}}}{V_{S_{Int}} + V_{S_{Dem}} - V_{S_{I}}}.
        \label{eq:pointingagreement2}
    \end{dmath}

\noindent \textsc{\textbf{User Preferences}}
We also conducted a post-test assessment employing a user preference questionnaire that participants answered after completing the outlining task under each condition.
We selected three sub-scales from the original social presence questionnaire proposed by Harms et al.~\cite{harms2004internal}. This questionnaire included 13 statements regarding co-presence (i.e., the feeling of being together in the same shared space), attentional allocation (i.e., amount of attention given to each other), and perceived message understanding, to be answered on a six-point Likert-scale, from Strongly Disagree (1) to Strongly Agree (6).
We also included two open questions at the end of the questionnaire: one that queried users if they were able to identify any strange behaviors ("Did you notice any strange behavior during the session?"), to evaluate if participants noticed the manipulation we were employing to their remote partner's avatar, and another for further comments/improvements about the techniques and the general user study ("Observations and Suggestions").

\subsection{Analysis}

\begin{figure}[t!]
\centering
\includegraphics[width=.95\columnwidth]{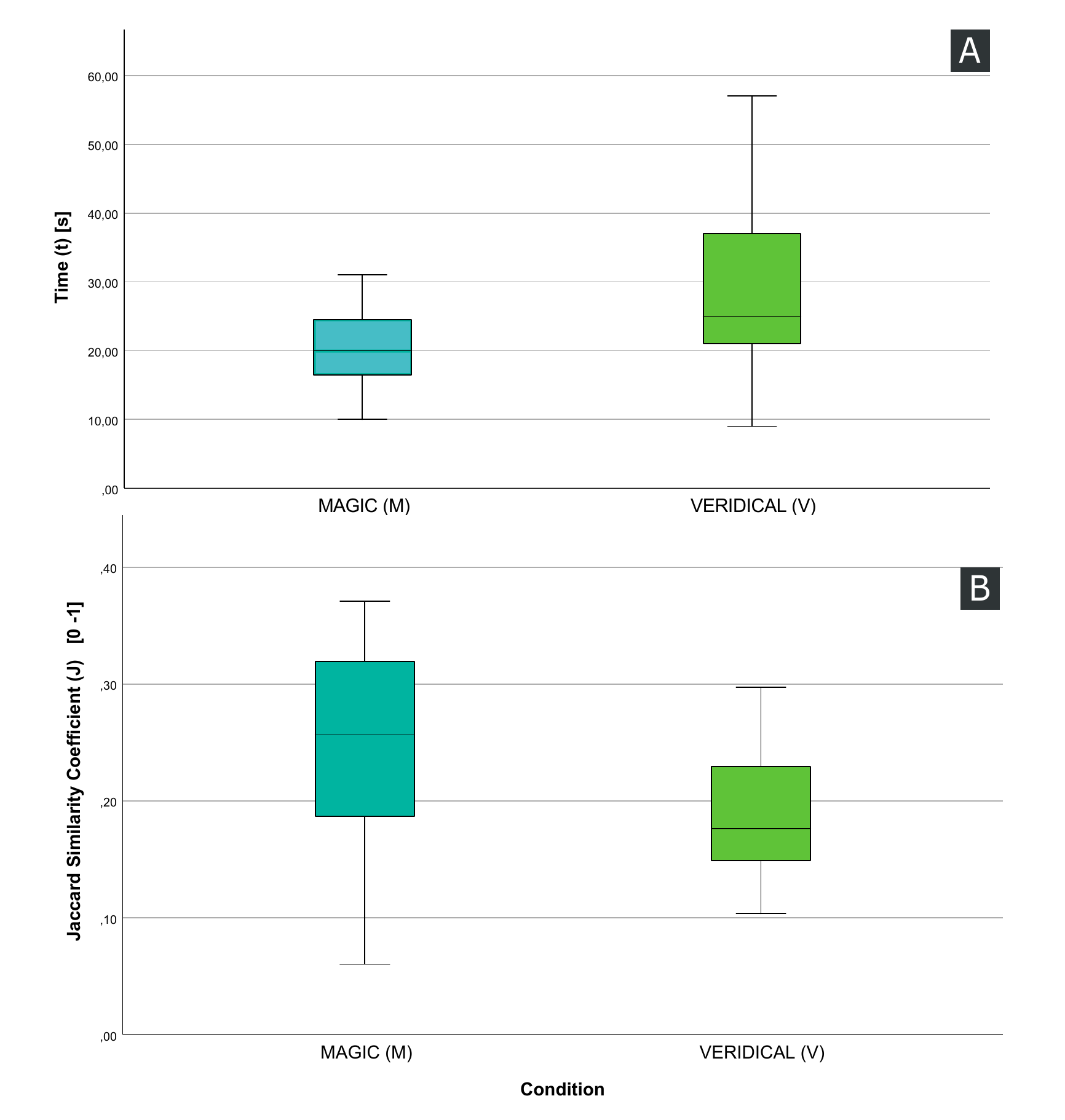}
\caption{Task performance results for each condition: A) Completion Time, \textit{t}.  B) Jaccard Similarity Coefficient, \textit{J}.}
\label{fig:results}
\end{figure}

\noindent \textsc{\textbf{Task Performance}}
A Shapiro-Wilk test showed that results for both metrics were normally distributed.
A paired t-test indicated that \textit{J} was statistically significantly higher for MAGIC (\textit{M} = 0.24, \textit{SD} = 0.03) than for the Veridical condition (\textit{M} = 0.18, \textit{SD} = 0.02), t(11) = -2,899, p = .014. 
As for task completion time, we found no statistically significant differences between conditions, t(11) = -2.00, p = .07.
Figure~\ref{fig:results} shows results for task performance metrics. 

\noindent \textsc{\textbf{User Preferences}}
\label{sec:UserPref}
In terms of 
co-presence, the Wilcoxon test reported a significant difference for Statement 1.2 (Z = -2.121, p = .034), revealing an increase in feeling of co-presence of the remote partner in the virtual environment under MAGIC.
Questionnaires showed no statistically significant differences between either approach in terms of Attentional Allocation and Perceived Message Understanding. 
However, we observed a tendency for better message understanding under the  MAGIC condition. 
Furthermore, when performing under MAGIC as Condition 2, three participants reported in the "Observations and Suggestions" open question that targets were easier to understand, and  one 
participant pointed out that they had to "move less than previously" to understand what their partner was pointing to. One participant also pointed that targets were harder to understand in the second part of the experiment when performing under Veridical as Condition 2. Details on the questionnaires and associated results can be found in the Appendix, Figure \ref{fig:table-user-preferences}.

\subsection{Discussion}



Results obtained from the statistical analysis of the collected data confirm our research hypothesis, \textbf{$H$}.
Results showed a higher pointing agreement for both participants when sharing perspective, yielding a better understanding of the collaboration tasks.  
Our approach improved pointing agreement by 6\%, representing a relative improvement of 33.3\% compared to the Veridical baseline condition . 
Participants also reported a better understanding of the target location under our approach, with some commenting that they had to move less to understand their partner's pointing gestures.
However, the questionnaire results did not show any statistically significant differences in Perceived Message Understanding statements between either condition. This could be due to a false sensation of agreement that was more prevalent under the Veridical condition, i.e., people thought they perceived their partner's actions correctly when in fact, they failed the outlining task. 
Therefore, pointing agreement dropped under the Veridical condition even though participants felt a similar personal performance.

Additionally, results showed no statistically significant differences in time completion under either condition. We hypothesise this to be due to a false sensation of agreement during the veridical condition that led participants to just move on to the next task even though they "failed" the latter.


As both conditions employed a life-sized representation of the remote avatar in a face-to-face configuration, we can say consequential communication was ensured under both conditions. 
Yet, when sharing perspective, there was an increase of feedthrough mechanisms, as the manipulations of artifacts present in the workspace were visible to the local observer at all times, contrary to what happened in the Veridical condition. 
The increased agreement on the pointing task also highlights enhanced intentional communication under MAGIC, as people showed a better capacity to use and understand pointing gestures. 
Hence, we argue that MAGIC generally improves workspace awareness in terms of feedthrough, consequential, and intentional communication mechanisms.

As an additional finding, we should note that answers from the User Preferences Questionnaire reported a significant difference for Statement 1.2, revealing participants felt that their partner was more present in the virtual environment under MAGIC. 
We argue that MAGIC's improvement in co-presence might be due to the fact that since the actions the remote collaborator performs in the workspace are both more visible and better understood, people feel their partner is more present in the workspace.

\revision{Regarding possible uncanny valley effects, our avatar was deliberately chosen as an abstract androgynous avatar, as these avatars are less likely to affect presence and embodiment significantly in unnatural behaviors \cite{uncanny}.}
\remove{Furthermore,} We directly asked participants in the User Preferences Questionnaire whether they noticed any "strange" behaviors when performing the tasks.
None of the participants reported noticing anything strange under MAGIC or the Veridical conditions, which leaves us to assume that the local collaborator did not notice distortions of the remote avatar, \revision{hence not revealing significant uncanny valley effects. However, we acknowledge that this could be due to the unrealistic avatar in itself, and that our distortions could potentially cause significant effects on presence and embodiment in more realistic avatars. To tackle this problem, the distortions we perform on the user’s movements could be optimized in the future by manipulating all points of a cloud representation instead of manipulating the three arm joints we used in our current implementation.}
\remove{Thus we can say that MAGIC is transparent to face-to-face collaborations.}

We also observed that participants performed the tasks using pointing gestures only and used verbal cues for keeping track of the partner's understanding of their performed actions (with sentences like "This is the area I want to show you," "Done!"), typically with one sentence before starting outlining and one sentence when finishing. As we developed our approach to tackling the problem of the complexity of trying to explain relative positions through words, considering pointing comes as such a natural behavior to humans that avoids confusion arising from speech, we assumed that if one can easily point and say “here,” one would not feel the necessity to use “to your left/ right,” eliminating communication errors related to handedness. Indeed, no participant used relative position words such as left and right during our experiments.

In general, MAGIC improves pointing agreement, thus increasing task performance.
Furthermore, MAGIC generally improves workspace awareness regarding feedthrough-consequential- and intentional- communication.
Additionally, MAGIC distortions of remote avatars are not noticeable or impairing. Finally, MAGIC positively impacts the presence of remote collaborators in the workspace. 
Thus, we can affirm that \textit{\say{a face-to-face setting coupled with manipulations that enable perspective sharing improves the understanding of pointing gestures, increasing workspace awareness.}}.


\subsection{Limitations}


Our experiments assumed static participant positions around the shared workspace.
Further research is needed to see whether these results scale to moving (or more than two) participants. 

\revision{Since our work falls into the category of a shared perspective \textit{with the twist of maintaining the advantages of a face-to-face formation}, we were interested in comparing this with traditional face-to-face formations. However, we acknowledge that a comparison with a traditional share of perspective could have also been interesting.}

\revision{Additionally, our approach was tested through an observational task, where users did not have the chance to manipulate the object.  We wanted to introduce our new paradigm of view sharing and test its feasibility in a simple case still representative of design review scenarios: pointing to regions of interest of a fixed 3D object for the discussion of relevant points.} 

\revision{Our approach conceptually guarantees that there is a share of the same perspective of the workspace for both users. In our use case and implementation, this workspace was static, and only the remote user’s movements were manipulated. We believe that applying similar manipulations to a dynamic workspace will guarantee that both users keep the same perspective of the workspace, hence generalizing to interactive scenarios.
We envision 
future work to include 
manipulation of 3D objects with MAGIC.}

We should also note that our approach doesn't avoid hand occlusions. Future work could address this problem by manipulating posture to ensure visibility at any time by, e.g., always bringing the hand to the top layer of rendering.

The experiments focused on objects commensurable with participants' apparent sizes. An open question remains on how to explore the trade-offs between apparent size and level of detail. 

Furthermore, MAGIC does not work with distal pointing since the manipulated pointing gestures only reference the correct place when the user touches the desired area. It could be interesting to extend our approach to enable distal pointing by dynamically changing the user representation depending on the task at hand~\cite{IterativelyAdaptingAvatars}, exploring different \textit{Beyond Real Interactions}\cite{BeyondReal} such as extending the user's arm in order to facilitate reaching an object as inspired by works such as in the Go-Go technique~\cite{GoGoTechnique}. 

Our technique does not ensure verbal communication between participants located in separate physical spaces, despite it being crucial in the course of the collaboration. 
During our experimental sessions, participants could talk to each other since they shared the same room.
Our findings suggest that co-localization only impacted verbal communication between participants, implying that our technique should be as effective for audio-enabled remote collaboration.

Additionally, during our studies, we asked pairs of participants to repeatedly point to the exact position of an object for internal validity, so that we could understand if MAGIC increases understanding of pointing gestures. A less controlled experiment will be valuable to evaluate further MAGIC's impact in a more general collaborative object-centered task. 

It is also worth mentioning that the experimental sessions were conducted during the pandemic which presented challenges in gathering participants, 12 total. Nevertheless, our results are statistically significant.

Finally, our results show an increase of \textit{J} from a medium value of 0.18 in the veridical condition to a medium value of 0.24 with MAGIC, representing an improvement of 33.3\%. These illustrate situations where the pointing agreement is close to what is depicted in Figure \ref{fig:PA_Scenarios} - 3rd left. Although this value has room for improvement, we believe this increase to be relevant in this context.

\section{Conclusions and Future Work}

We introduced MAGIC, an approach to improve remote collaboration by enhancing pointing agreement in object-centered collaboration in virtual environments. 
Our approach allows people to be in a face-to-face f-formation, promoting the sense of co-presence, and improving object-related communication accuracy.  
MAGIC addresses the problem of different viewpoints and the natural occlusions that arise from it, by allowing people to view the same workspace PoV while subtly altering the remote participant gestures to match the same reference space. 
Our work is the first to support face-to-face collaboration with a shared viewpoint without apparent avatar manipulation or resorting to additional tools to perform the communication (e.g., pointers).

Results from a user study show that MAGIC improved participants' pointing agreement during 
tasks using pointing gestures and increased the sense of co-presence, 
while being unnoticeable to 
users.
These findings validate our initial hypothesis and clear the way for future research on remote face-to-face meetings 
in virtual environments.
Although our study was conducted in VR, we envision our approach to be relevant for general Mixed Reality scenarios. However, this generalization to MR was not formally evaluated, remaining as future work.

Future work could extend our approach to multi-user settings and apply MAGIC to more complex real-life tasks\revision{, expanding to dynamic workspaces where users can manipulate the objects under analysis}.
Moreover, as we could see that subtle manipulations of pointing gestures had such positive impacts, future work should focus on more complex body representations.  
Interestingly, gaze and posture can communicate different intents in non-verbal ways, including cooperativeness, social status, and turn-taking, among others~\cite{holograms19}. We would be interested in exploring whether we can manipulate such cues by subtly warping a remote speaker's gaze and posture (e.g., correcting one's gaze and posture to promote likeness and cooperation).
\revision{Additionally, while other works have explored other ways of improving the use of pointing gestures in collaborative environments by complementing these with the use of pointers and sketches \cite{kim2019,DBLP:conf/iccsa/ConteroNJC03,pereira2004cascading}, we were interested in improving “raw pointing” to mimic real-world interactions better. Whether our approach could be further improved by complementing it with external tools remains future work.}
Finally, we designed our approach to manipulate remote user movements imperceptibly. 
However, we would like to explore whether perceptible exaggerated distortions have the potential to improve collaboration. 
Many inspiring examples come from the \textit{Commedia dell'Arte}  
soap operas, cartoons, mime, and many other forms of communication through exaggeration, and more generally looking at user interfaces as theatre~\cite{laurel14}.
\acknowledgments{
This work is financed by \textit{Fundação para a Ciência e a Tecnologia} (Portuguese Foundation for Science and Technology) through grants 2022.09212.PTDC (XAVIER), UIDB/50021/2020 and Carnegie Mellon Portugal grant SFRH/BD/151465/2021 under the auspices of the UNESCO Chair on AI\&XR.
}

\bibliographystyle{abbrv}

\bibliography{acmart}

\pagebreak

\onecolumn
\section*{APPENDIX}

\begin{table}[h]
\centering
\includegraphics[width=.95\textwidth]{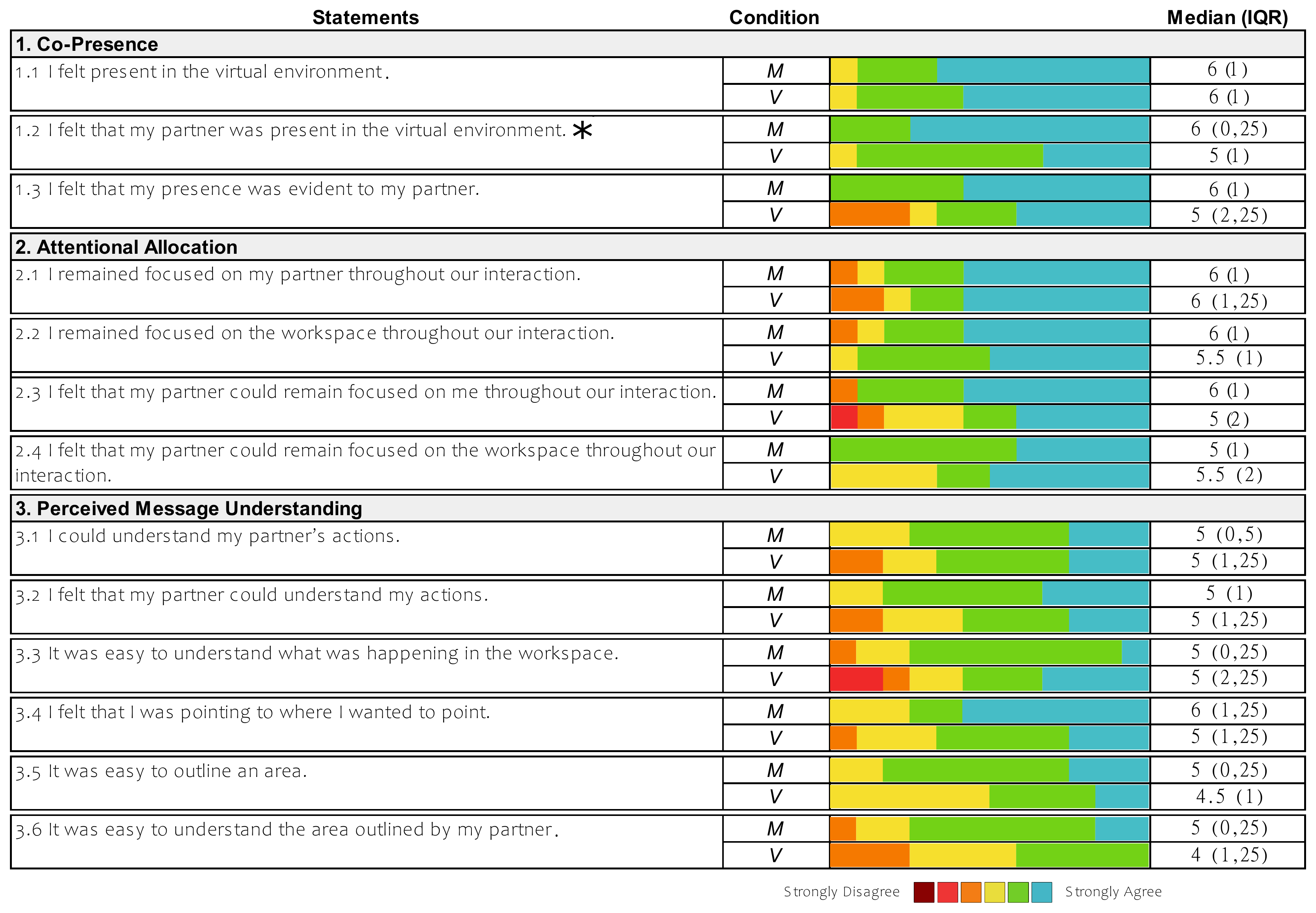}
\caption{Results from the user preferences questionnaire for both MAGIC(M) and Veridical(V) conditions. * indicates statistical significance.}
    \label{fig:table-user-preferences}
\end{table}


\end{document}